\documentclass[12pt,a4paper]{article}

\newcommand{\ppbar}{p^{\!\!\!\!\!\textsuperscript{\tiny{(--)}}}\!\!}
\newcommand{\ptmiss}{{p\!\!\!\!\! \not \,\,\,\,}_t}
\providecommand{\openone}{\leavevmode\hbox{\small1\kern-3.8pt\normalsize1}}

\usepackage{amssymb,graphicx}
\usepackage{epsfig}
\usepackage{cite}

\begin{document}

\begin{center}
\begin{Large}
{\bf Neutrino physics at large colliders}

\end{Large}

\vspace{0.5cm}
F. del Aguila$^1$, J. A. Aguilar--Saavedra$^1$, R. Pittau$^2$  \\[0.2cm] 
{\it $^1$ Departamento de Física Teórica y del Cosmos and CAFPE, \\
Universidad de Granada, E-18071 Granada, Spain} \\[0.1cm]
{\it $^2$ Dipartimento di Fisica Teorica, Universit\`a di Torino, 
and INFN \\ Sezione di Torino, Italy} 
\end{center}

\begin{abstract}
Large colliders are not sensitive to light neutrino masses and character, but 
they can produce new heavy neutrinos, allowing also for the determination
of their Dirac or Majorana nature. We review
the discovery limits at the next generation of large colliders.
\end{abstract}

\section{Introduction}
\label{Introduction}

Future colliders will probe Nature up to TeV scales with high precision,
discovering new heavy particles  with sizeable couplings or setting stringent
limits  on their existence. Thus, we expect that the Higgs will be copiously
produced at the LHC, as may also be the lightest superpartners
\cite{atlas,cms}. At any rate, these facilities will be a window to any new
physics near the electroweak scale which couples to the Standard Model (SM). 
This could be the case, for example, of new heavy neutrinos. In the following
we review what can be learned about them at the next generation of large
colliders. 

SM extensions with heavy neutrinos at the TeV scale or below usually include new
interactions and extra matter. If the latter are also in this mass range, they
give new signals and further contributions to heavy neutrino production
processes. We neglect them here, assuming that they do not affect the
significance of the specific processes discussed. (In this sense the resulting
limits are conservative, because we do not consider other possible effects
eventually larger.) Hence, all the additional parameters involved in
heavy neutrino production at colliders are the heavy neutrino
masses and their mixings with SM fermions. Let $N$ be the lightest heavy
neutrino, which can be of Dirac (D) or Majorana (M) nature. In the former case
two  extra bispinors $N_{L,R}$ are added and lepton number is conserved, whereas
in the latter the SM is enlarged with only one bispinor satisfying
$(N_L)^c \equiv C \bar N_L^T = N_R$, and thus lepton number is violated
\cite{MP,BLS}. It should be stressed that: 

\begin{itemize}
\item The effects of light neutrino masses at large colliders are suppressed by
factors $\sim m_\nu / \sqrt{s}$, and they can be completely neglected. Hence, we
can assume without loss of  generality that the SM neutrinos are strictly
massless, and lepton number is preserved in the light sector. In other words, 
it does not matter in this context whether the light neutrinos are of Dirac or
Majorana nature \cite{Bilenky}.\footnote{In definite models with extra
interactions and matter content light neutrino masses and mixing parameters may
be related to the new sectors and, through them, to collider observables (see
for examples Refs. \cite{porod1,porod2}). Then, the latter
can give indirect information about the former, which are otherwise
unmeasurable at high energies.}
\item The heavy Dirac neutrino case is a particular limit of two heavy Majorana
neutrinos, when both are degenerate and their mixings with the SM leptons are
lepton number conserving (LNC). On the other hand, one extra heavy Majorana 
neutrino corresponds to the limit in which the second Majorana neutrino is much
heavier. 
\end{itemize}

In both situations, the relevant additional terms of the Lagrangian are the
heavy neutrino mass term 
\begin{eqnarray}
\mathcal{L}_\mathrm{mass} & = & - m_N \bar N_L N_R  + \mathrm{H.c.} \,,
  \quad \quad ~~\, (\mathrm{D})  \nonumber \\ 
\mathcal{L}_\mathrm{mass} & = & - \frac{1}{2} \, m_N \bar N_L N_R +
 \mathrm{H.c.}  \,, \quad \quad (\mathrm{M}) 
\label{ec:massterm} 
\end{eqnarray}
(with $N_{L,R} = P_{L,R} N$), and its interactions
with
the light fermions \cite{clic}. The charged current vertex with a
charged lepton $\ell$ reads 
\begin{equation}
\mathcal{L}_W = - \frac{g}{\sqrt 2}  \left( \bar \ell \gamma^\mu V_{\ell N}
P_L N \; W_\mu + \bar N \gamma^\mu V_{\ell N}^* P_L \ell \; W_\mu^\dagger
\right) \,.
\label{ec:lNW} 
\end{equation}
The neutral current gauge couplings with a light neutrino $\nu_\ell$ are
\begin{equation}
\mathcal{L}_Z = - \frac{g}{2 c_W}  \left( \bar \nu_\ell \gamma^\mu
V_{\ell N} P_L N + \bar N \gamma^\mu V_{\ell N}^* P_L \nu_\ell \right)
Z_\mu \,,
\label{ec:nNZ}
\end{equation}
and the scalar interactions
\begin{equation}
\mathcal{L}_H = - \frac{g \, m_N}{2 M_W} \, \left( \bar \nu_\ell \, V_{\ell N}
P_R N + \bar N \, V_{\ell N}^* P_L \nu_\ell \right) H \,.
\label{ec:nNH}
\end{equation}
Thus, all the necessary parameters for the evaluation of the $N$ production
cross
sections are its mass $m_N$ and its couplings to the different lepton flavours 
$V_{\ell N}$, $\ell = e, \mu, \tau$. The main difference between a heavy Dirac
or Majorana neutrino is the non-zero lepton number violating (LNV) propagator
for two $N$ fields $ \langle N N \rangle$ in the Majorana case, besides the
common LNC one $\langle N\bar N \rangle$ \cite{Denner}.
Dirac and Majorana heavy neutrinos have the same couplings but
for the Majorana case the second term in Eqs.~(\ref{ec:nNZ}) and 
(\ref{ec:nNH}) can be rewritten using the condition  $N = N^c$,
giving\footnote{This also assumes that $\nu=\nu^c$, otherwise both fields $\nu$
and $\nu^c$ would appear. As it has been emphasised, results are independent of
this assumption.} 
\begin{eqnarray}
\mathcal{L}_Z & = & - \frac{g}{2 c_W} \, \bar \nu_\ell \gamma^\mu \left(  
V_{\ell N} P_L - V_{\ell N}^* P_R \right) N \; Z_\mu \,, \quad \quad
\mathrm{(M)} \nonumber \\
\mathcal{L}_H & = & - \frac{g \, m_N}{2 M_W} \, \bar \nu_\ell \left( V_{\ell N}
 P_R + V_{\ell N}^* P_L \right) N \; H \,. \quad \quad \quad
\mathrm{(M)}
\label{ec:nNZHm}
\end{eqnarray}
For equal values of the couplings, production cross sections are equal in
general for a Dirac and Majorana heavy neutrino, as we will explicitly see
in section \ref{sec:5} (in some processes, this requires summing $N$ and
$\bar N$ production for Dirac neutrinos). 
For the analysis of the different heavy neutrino signals it is necessary to
know its decay modes as well. $N$ can decay in the channels $N \to W^+ \ell^-$ 
(if $N$ is a Majorana fermion $N \to W^- \ell^+ $ is also allowed), 
$N \to Z \nu_\ell$ and $N \to H \nu_\ell$. 
The partial widths for these decays are \cite{clic,GZ1,Pil3}
\begin{eqnarray}
\Gamma(N \to W^+ \ell^-) & = &  \Gamma(N \to W^- \ell^+) \nonumber \\
& = & \frac{g^2}{64 \pi} |V_{\ell N}|^2
\frac{m_N^3}{M_W^2} \left( 1- \frac{M_W^2}{m_N^2} \right) 
\left( 1 + \frac{M_W^2}{m_N^2} - 2 \frac{M_W^4}{m_N^4} \right) \,, \nonumber
\\[0.1cm]
\Gamma_D(N \to Z \nu_\ell) & = &  \frac{g^2}{128 \pi c_W^2} |V_{\ell N}|^2
\frac{m_N^3}{M_Z^2} \left( 1- \frac{M_Z^2}{m_N^2} \right) 
\left( 1 + \frac{M_Z^2}{m_N^2} - 2 \frac{M_Z^4}{m_N^4} \right) \,, \nonumber
\\[0.2cm]
\Gamma_M(N \to Z \nu_\ell) & = & 2 \, \Gamma_D(N \to Z \nu_\ell) \,, \nonumber
\\
\Gamma_D(N \to H \nu_\ell) & = &  \frac{g^2}{128 \pi} |V_{\ell N}|^2
\frac{m_N^3}{M_W^2} \left( 1- \frac{M_H^2}{m_N^2} \right)^2 \,, \nonumber
\\[0.2cm]
\Gamma_M(N \to H \nu_\ell) & = & 2 \, \Gamma_D(N \to H \nu_\ell) \,,
\label{ec:widths}
\end{eqnarray}
with $\Gamma_D$ and $\Gamma_M$ standing for the widths of a Dirac and Majorana
neutrino. The factors of two in the partial widths of $N \to Z \nu_\ell$,
$N \to H \nu_\ell$ for a Majorana neutrino are the consequence of the extra
$V_{\ell N}^*$ couplings in Eqs.~(\ref{ec:nNZHm}) resulting from the Majorana
condition, and which are not present for a Dirac neutrino. From
Eqs.~(\ref{ec:widths}) it follows that for equal values of the mixing angles
$V_{\ell N}$ the width of a heavy Majorana neutrino is twice as large as for a
Dirac neutrino.
Another straightforward consequence of the previous expressions is that 
the partial widths for $W$, $Z$ and Higgs decays are in the ratios 
$2 \, : \, 1 \, : \, 1$ (assuming $m_N \gg M_W, M_Z, M_H $). 

The reader mainly interested in the reach of the different experiments 
can go directly to section \ref{sec:5}, where we
review the results for heavy neutrino production at future colliders. In
next section we discuss the motivation for having heavy neutrinos at
the electroweak scale, and in section \ref{sec:3} we provide more 
details on the neutrino mass matrices we are interested in, 
for easier comparison with other analyses. 
Present indirect constraints on heavy neutrino mixing are summarised in
section \ref{sec:4}. Section \ref{sec:6} is devoted to our conclusions.

\section{Motivation}
\label{sec:2}

New fermions are present in many extensions of the SM. If their masses are well
above the electroweak scale they must be vector-like, that is, with their
left- and right-handed parts transforming in the same represetation of the SM
gauge group.  This guarantees that their mass terms preserve the electroweak
gauge symmetry,\footnote{In particular, fermions transforming trivially under
the SM group can have Majorana nature. In this case particles and antiparticles
coincide, and their (Majorana) masses violate fermion number.}
but by the same token the natural size of their masses is of the order of the
largest mass scale allowed by the symmetries of the model \cite{thooft}.
In grand unified
theories this is usually the grand unified  scale, near the Planck mass. 
This is for example the case of Majorana right-handed neutrinos
\cite{seesaw1,seesaw2,seesaw3,seesaw4}.
However, in
the presence of extra symmetries these vector-like masses can be much smaller
\cite{ramond,paco,LL}, and exotic fermions can also manifest at lower
energies (for reviews see Refs.~\cite{frampton,hewett}).

This scenario has become more interesting lately because the possible existence
of low extra dimensions near the TeV scale \cite{extra1,extra2}, with SM
fermions in the bulk, implies infinite towers of vector-like fermions, known as
Kaluza-Klein modes \cite{torre1,torre2,torre3},
whose lightest states could be also
eventually observable \cite{jose,gustavo}. 
Fermions in five dimensions have both chiralities and, although the zero modes
can be chiral in appropriate backgrounds, the Kaluza-Klein modes are in general
Dirac particles. If neutral, they can have Majorana mass terms as well
\cite{a31,a32}.
Moreover, as an alternative to supersymmetry at the TeV scale, SM extensions
with a larger global symmetry have  been proposed to cancel the undesired
large quantum corrections to the Higgs mass \cite{LH1,LH2,LH3}.
In these models the SM
Higgs is a pseudo-Goldstone  boson, what justifies their name of Little Higgs
models.  They are characterised by a new scale near the TeV and include
extra matter
content to realise the larger symmetry. In this context the existence of new
fermions near the  electroweak scale is not a possibility but a
requirement. In general they mix with the SM fermions and can be 
eventually produced at large colliders \cite{coll1,coll2,coll3}.
In particular, if the model includes extra neutrinos transforming as singlets
under the SM group, it seems natural that they have TeV masses and relatively
large mixings with the SM fermions.
For example, in the {\em simplest} Little Higgs models \cite{schmaltz},
the matter
content belongs to $\mathrm{SU}(3)$ multiplets, and the SM lepton doublets must
be enlarged with one extra neutrino $N'_{\ell L}$ per family. When a combination
of Higgs triplets acquires a large vacuum expectation value (VEV) $f$ of few
TeV, reducing
the $\mathrm{SU}(3)$ symmetry down to $\mathrm{SU}(2)_L$, the extra neutrinos
can get a large Dirac mass of the order of the new scale, provided the model
also includes the necessary right-handed neutrinos transforming as
$\mathrm{SU}(3)$ singlets \cite{padilla}. This mechanism provides a natural way
of giving masses to the SM neutrinos while explaining
the absence of exotic fermions below the electroweak scale. In this framework
the mixing between the light leptons and the heavy neutrinos is of order
$v/\sqrt 2 f$, with $v=246$ GeV the electroweak VEV.
  If this is the case, it is
precisely the non-observation  of these fermions at future colliders what will
set very  stringent limits on the model.

Besides their appearance in specific models, 
heavy Majorana neutrinos are introduced in seesaw models
\cite{seesaw1,seesaw2,seesaw3,seesaw4}. They
give contributions to light neutrino masses $m_\nu$ of the order
$Y^2 v^2  / 2 m_N$, where $Y$ is a Yukawa coupling.
In the minimal seesaw realisations
this is the only source for
light neutrino masses, and the Yukawas are assumed of order unity without any
particular symmetry. Therefore,
having $m_\nu \sim Y^2 v^2 / 2 m_N$ requires heavy masses $m_N \sim 10^{13}$ GeV
to reproduce the observed light neutrino spectrum. Additionally, the light-heavy
mixing is predicted to be $V_{\ell N} \sim \sqrt{m_\nu / m_N}$.
These ultra-heavy particles are unobservable, and thus the seesaw mechanism
is not directly testable. Nevertheless, non-minimal
seesaw models can be built, with $m_N \sim 1$ TeV or smaller, if some 
approximate flavour
symmetry suppresses the $\sim  Y^2 v^2 / 2 m_N$ contribution from seesaw
\cite{buch2,ing,gluzatev}.
Moreover, the light-heavy mixing can be decoupled from the mass ratio
$\sqrt{m_\nu / m_N}$ \cite{bernabeu}. In this situation, the heavy states could
be observable in future collider experiments.

In addition to providing a mechanism for neutrino mass generation,
heavy neutrino decays can give a succesful leptogenesis. A beautiful feature of
the minimal seesaw
is that the same heavy neutrino scale $m_N \sim 10^{13}$ GeV which reproduces
light neutrino masses predicts a lepton asymmetry large enough to account for
the observed baryon asymmetry through $B+L$ violating sphaleron interactions 
\cite{lepto1,lepto2}.
The heavy neutrino scale can be lowered to the TeV, if two neutrinos are
nearly degenerate and the CP asymmetry is resonantly enhanced
\cite{qd1,qd2,qd3}, or for other SM extensions \cite{tev1,tev2,RH}.
It seems to be a general property that the heavy neutrino
needed to generate the CP asymmetry must couple very weakly to the SM
fields, and then it cannot be produced at existing or planned
colliders through the usual mechanisms.
But this does not preclude the existence of other heavy neutrinos with larger
couplings which, although not participating actively in leptogenesis, may 
be observable at colliders. In the example of Refs.~\cite{tau,tau2}, with 3
heavy
neutrinos with masses $m_{N_i} \sim 250$ GeV, the observed light neutrino
spectrum is reproduced and a baryon excess is generated through a $\tau$ lepton
asymmetry.
The heavy neutrino $N_1$ actively involved in $\tau$ leptogenesis couples very
weakly to the SM leptons, $V_{\ell N_1} \sim 10^{-6}$, and cannot be produced
at existing or planned colliders. The other two heavy neutrinos $N_{2,3}$ 
form a quasi-Dirac neutrino $N$ mainly coupling to the first two families,
$V_{eN}, V_{\mu N} \sim 10^{-2}$, and with a small mixing with the third one,
$V_{\tau N} \sim 10^{-6}$, to avoid
washing-out the $\tau$ lepton asymmetry.
This heavy state could be observable at $e^+ e^-$ colliders.

\section{Beyond the Neutrino Standard Model}
\label{sec:3}

In this section we consider a SM extension with heavy neutrino singlets,
deriving their interactions with the light leptons \cite{valle}. The general
situation is that $n$ additional singlets $N'_{iL}$ ($i=1, ..., n$),
which can be taken to be left-handed, are introduced.\footnote{In
the case of light Dirac neutrinos we must introduce 3 additional
$\mathrm{SU}(2)_L$ singlet fields to allow for Dirac neutrino masses. However,
as it has already been emphasised, light neutrino masses are not relevant at
large colliders, so these fermions can be considered massless and their
right-handed parts ignored. Any other more general scenario can be brought into
this form in this limit. In particular, this is consistent with whatever
specific model of neutrino masses one may advocate. Here we do not make any
attempt to accommodate them, neither claim that the heavy neutrino masses and
mixings we consider are natural in any given model of neutrino mass generation
one can think of. But heavy neutrinos can be introduced, as long as they 
fulfill the experimental constraints discussed in the next section and can
reproduce the 
pattern of neutrino masses and mixings implied by neutrino oscillations.}
We will assume $n = 3$ for definiteness and easier comparison with 
the heavy neutrino production analysis we are interested in,
but the formalism is general for any number of singlets. 
The neutrino weak isospin $T_3=1/2$ eigenstates
$\nu'_{iL}$ are the same as in the SM, and the extended mass term reads
\begin{eqnarray}
\mathcal{L}_\mathrm{mass} & = & - \frac{1}{2} \,
\left(\bar \nu'_L \; \bar N'_{L} \right)
\left( \! \begin{array}{cc}
M_L & \frac{v}{\sqrt 2} Y \\ 
\frac{v}{\sqrt 2} Y^T & M_R 
\end{array} \! \right) \,
\left( \!\! \begin{array}{c} \nu'_R \\ N'_{R} 
\end{array} \!\! \right)
\; + \mathrm{H.c.} \,,
\label{massterm}
\end{eqnarray}
where 
$\nu'_{iR} \equiv (\nu'_{iL})^c$, $N'_{jR} \equiv (N'_{jL})^c$, and the blocks
$M_L$, $Y$ and $M_R$ stand for $3 \times 3$ matrices. $M_L$ is a lepton number
violating mass matrix for the light neutrinos, $Y$ stands for the Yukawa
interactions and $M_R$, which can be assumed diagonal, real and positive
without loss of generality, corresponds to bare masses which also violate
lepton number. The mass eigenstates are obtained diagonalising 
the complete $6\times 6$ mass matrix $\mathcal{M}$ with the  
unitary transformation $\mathcal{U}_L$,
\begin{equation}
\left( \!\! \begin{array}{c} \nu_L \\ N_L \end{array} \!\! \right) =
\mathcal{U}_L^\dagger
\left( \!\! \begin{array}{c} \nu'_L \\ N'_L \end{array} \!\! \right) \,,
\quad \quad 
\left( \!\! \begin{array}{c} \nu_R \\ N_R \end{array} \!\! \right) =
\mathcal{U}_L^T
\left( \!\! \begin{array}{c} \nu'_R \\ N'_R \end{array} \!\! \right) \,,
\label{ec:U}
\end{equation}
with $\mathcal{U}_L^\dagger 
\mathcal{M} \, \mathcal{U}_L^* = \mathcal{M}_\mathrm{diag}$.
The weak interaction Lagrangian in the weak eigenstate basis is
\begin{eqnarray}
\mathcal{L}_W & = & - \frac{g}{\sqrt 2} \, \bar l'_L \gamma^\mu \nu'_L
W_\mu + \mathrm{H.c.} \,, \nonumber \\
\mathcal{L}_Z & = 
& - \frac{g}{2 c_W} \, \bar \nu'_L \gamma^\mu \nu'_L Z_\mu 
\,, \nonumber \\
\mathcal{L}_H & = & - \frac {1}{\sqrt 2} \; \bar \nu'_L Y N'_R \, H  
+ \mathrm{H.c.} \, ,\end{eqnarray}
with $l'_{iL}$ the charged lepton weak eigenstates. Using Eqs.~(\ref{ec:U}),
these terms read in the mass eigenstate basis
\begin{eqnarray}
\mathcal{L}_W & = & - \frac{g}{\sqrt 2} \, \bar l_L \gamma^\mu \; 
U_l^\dagger U_L
\left( \!\! \begin{array}{c} \nu_L \\ N_L \end{array} \!\! \right) 
W_\mu + \mathrm{H.c.} \,, \nonumber \\
\mathcal{L}_Z & = & - \frac{g}{2 c_W} 
\left(\bar \nu_L \; \bar N_L \right)
\gamma^\mu \; U_L^\dagger U_L
\left( \!\! \begin{array}{c} \nu_L \\ N_L \end{array} \!\! \right)
Z_\mu \,, \nonumber \\
\mathcal{L}_H & = & - \frac {1}{\sqrt 2}  \left(\bar \nu_L \; \bar N_L \right)
U_L^\dagger Y U_L^{'*}
\left( \!\! \begin{array}{c} \nu_R \\ N_R \end{array} \!\! \right)
 \, H  + \mathrm{H.c.}  \, ,
\label{ec:wzh}
\end{eqnarray}
where the $3 \times 6$ matrices $U_L$ and $U'_L$ 
are the upper and lower blocks of $\mathcal{U}_L$, 
\begin{equation}
\mathcal{U}_L = \left( \! \begin{array}{c} U_L \\ U'_L \end{array} \! \right)
\; , 
\label{mixingm}
\end{equation}
and $U_l$ is a $3 \times 3$ unitary matrix resulting from the 
diagonalisation of the charged lepton mass matrix.
In the weak basis where the latter is diagonal 
$U_l$ is the identity, so we can assume without loss of generality   
that the extended $3 \times 6$ Maki-Nakagawa-Sakata (MNS) matrix
\cite{PMNS1,PMNS2}
\begin{equation}
V \equiv U_l^\dagger U_L
\end{equation}
equals $U_L$. Hence, from
Eqs.~(\ref{ec:wzh}) we see that this matrix completely fixes the couplings 
of light leptons to heavy neutrinos, including their scalar 
interactions, although in this case the couplings are also proportional to the
heavy neutrino masses $m_{N_i}$. The charged current matrix can be further
decomposed into $3 \times 3$ blocks for convenience,
\begin{equation}
V = \left( V^{(\nu)} ~ V^{(N)} \right) \,,
\end{equation}
where $V^{(\nu)}$ and $V^{(N)}$ mix the charged leptons with light and heavy
neutrinos, respectively.
Its approximate form can be obtained
observing that the $3 \times 3$ blocks of the mass matrix $\mathcal{M}$ exhibit
a strong hierarchy $M_L \ll v Y \ll M_R$, where here $M_L$, $Y$ and $M_R$
stand for generic values of their entries.
Then, to first order in small ratios,
\begin{eqnarray}
\mathcal{U}_L & = &
\left( \! \begin{array}{c} U_L \\ U'_L \end{array} \! \right)
\simeq 
\left( \! \begin{array}{cc}
\openone & V^{(N)} \\ 
- V^{(N)\dagger} & \openone 
\end{array} \! \right) \,,
\label{mixingDa}
\end{eqnarray}
where $V^{(N)}$ has matrix elements 
\begin{eqnarray}
V_{\ell N_i} = \frac{v}{\sqrt 2}\frac{Y_{\ell i}}{M_i} \,,
\label{mixing}
\end{eqnarray}
being $M_i$, $i=1,\dots,3$ the eigenvalues of the diagonal matrix $M_R$.
Heavy neutrino masses are $m_{N_i} \simeq M_i$. Their couplings $V_{\ell N_i}$
are very small and it makes sense to keep only the first order terms.
(Experimental data constrain these quantities to be $O(10^{-2})$ at most,
see next section.)
$V^{(\nu)}$ is unitary up to small corrections $O(V_{\ell N_i}^2)$ and
has been taken as the $3 \times 3$ identity matrix,
because light neutrino masses can be safely neglected at high energies. 
The interactions for heavy Majorana neutrinos presented in the introduction
can be easily obtained substituting the expressions
for $\mathcal{U}_L$ and $V$ into Eqs. (\ref{ec:wzh}).
We note that while the charged current interaction has the form shown in
Eq.~(\ref{ec:lNW}) by definition, the neutral current and Higgs terms
correspond to the leading term for small $V^{(N)}$.
In the general case of three heavy Majorana neutrinos the vertices in
Eqs. (\ref{ec:lNW})--(\ref{ec:nNH}) correspond to any of them.

For a heavy Dirac neutrino the interactions are the same but some redefinitions
are necessary. In this case there are two degenerate eigenstates, say
$N_1$ and $N_2$, and their Yukawa couplings (the first two columns) are
proportional,
\begin{equation}
m_{N_2} = m_{N_1} \,, \quad  Y_{\ell 2} = i Y_{\ell 1} \,.
\label{Dirac}
\end{equation}
From the latter equality, $V_{\ell N_2} = i V_{\ell N_1}$.
These conditions realise the lepton number symmetry $L$ inherent to this case.
Defining
\begin{eqnarray}
N_L & \equiv & \frac{1}{\sqrt 2} (N_{1L} + i N_{2L}) \,, \nonumber \\
(N_R)^c & \equiv & \frac{1}{\sqrt 2} (N_{1L} - i N_{2L}) \,,
\end{eqnarray}
it is easily found that $N_L$ couples to the charged leptons with
\begin{equation}
V_{\ell N} = \sqrt 2 V_{\ell N_1} \,,
\end{equation}
while $(N_R)^c$ decouples. 
Lepton numbers $L=1$, $L=-1$ can be assigned to $N_L$, $(N_R)^c$, respectively.
Rewriting the Majorana mass term for $N_1$ and $N_2$
\begin{equation}
\mathcal{L}_\mathrm{mass} = - \frac{m_{N_1}}{2} \left (\bar N_{1L} (N_{1L})^c +
\bar N_{2L} (N_{2L})^c \right) + \mathrm{H.c.}
\end{equation}
gives the usual LNC mass term of a Dirac neutrino with
mass $m_N \equiv m_{N_1}$ in Eq.~(\ref{ec:massterm}),
\begin{equation}
\mathcal{L}_\mathrm{mass} = - m_N \bar N_L N_R
+ \mathrm{H.c.}
\end{equation}

\section{Experimental constraints}
\label{sec:4}

The mixing of neutrinos heavier than the $Z$ boson with charged leptons 
is limited by two sets of processes 
\cite{LL,pil1,nardi,pil2,bernabeu,kagan,illana}:
({\em i\/}) $\pi \to \ell \nu$, $Z \to \nu \bar \nu$ and other tree-level
processes involving light neutrinos in the final state; ({\em ii\/})
$\mu \to e \gamma$, $Z \to \ell^+ \ell^{'-}$ and other lepton flavour violating
(LFV) processes to which heavy neutrinos can contribute at one loop level.
These imits are independent of the Dirac or Majorana nature of the
neutrinos, and constrain the quantities
\begin{equation}
\Omega_{\ell \ell'} \equiv \delta_{\ell \ell'} - \sum_{i=1}^3 
V_{\ell \nu_i} V_{\ell' \nu_i}^* =  
\sum_{i=1}^3 V_{\ell N_i} V_{\ell' N_i}^* \,,
\label{ec:omega}
\end{equation}
where the equality is a consequence of the unitarity of 
$\mathcal{U}_L$. A global fit to the
processes in the first group gives the bounds \cite{kagan}
\begin{equation}
\Omega_{ee} \leq 0.0054 \,, \quad \Omega_{\mu \mu} \leq 0.0096 \,, \quad
\Omega_{\tau \tau} \leq 0.016 \,,
\label{eps1}
\end{equation}
with a 90\% confidence level (CL). In the limit of heavy neutrino masses 
in the TeV range, LFV processes in the second group require \cite{bernabeu}
\begin{equation}
|\Omega_{e \mu}| \leq 0.0001 \,, \quad |\Omega_{e \tau}| \leq 0.01 \,, \quad
|\Omega_{\mu \tau}| \leq 0.01 \,.
\label{eps2}
\end{equation}
The bounds in Eqs.~(\ref{eps1}) are model-independent to a large extent, and
independent of heavy neutrino masses as well. They imply that the mixing of the
heavy eigenstates with the charged leptons is very small, 
$\sum_i |V_{\ell N_i}|^2 \leq 0.0054$, 0.0096, 0.016 for $\ell = e,\mu,\tau$, 
respectively. 
On the other hand, in general the bounds in Eqs.~(\ref{eps2}) do not directly 
constrain the products $V_{\ell N_i} V_{\ell' N_i}^*$, but their sums,
and cancellations might occur between two or more terms 
and also with other new physics contributions. 
These cancellations may be more or less natural, but in any case
such possibility makes the limits in Eqs.~(\ref{eps2}) relatively
weak if more than one heavy neutrino exists \cite{plb,GZ1}. 
When discussing the future 
collider limits in next section we will take into account only the bounds 
in Eqs.~(\ref{eps1}), eventually using those in Eqs.~(\ref{eps2}) for
comparison.

Specific neutrino mass models must reproduce the observed light neutrino
spectrum.
In the block-diagonal basis defined by Eq.~(\ref{mixingDa})
the neutrino mass matrix in Eq.~(\ref{massterm})
reads (neglecting small corrections) 
\begin{eqnarray}
\mathcal{U}_L^\dagger \, \mathcal{M} \, \mathcal{U}_L^* & \simeq & 
\left( \! \begin{array}{cc}
M_L - \frac{v^2}{2}Y M_R^{-1} Y^T & 0 \\ 
0 & M_R 
\end{array} \! \right) \,. 
\end{eqnarray}
The light neutrino masses and mixings result from the diagonalisation of the
light neutrino $3 \times 3$ mass matrix
\begin{equation}
M_\nu \simeq M_L - \frac{v^2}{2}Y M_R^{-1} Y^T \,.
\end{equation}
The first term $M_L$ vanishes unless a Higgs triplet is included in the theory,
in which case it must be justified why its matrix elements are small.
For $V_{\ell N} \sim 0.01$ and $m_N \sim 1$ TeV, the typical size of the second
term is $0.1$ GeV, 8 orders of magnitude larger than the light neutrino mass
scale $m_\nu \sim 1$ eV. Then, one has to arrange either (i) some supression
mechanism, or (ii) a fine-tuned
cancellation with $M_L$. In any of these cases one expects that some
symmetry is at work. A particular example of the former appears when two
degenerate Majorana neutrinos form a (quasi) Dirac neutrino. In such case,
their contributions cancel due to the conditions in Eqs.~(\ref{Dirac}).
(If all heavy neutrinos are Dirac particles the light neutrinos can be also
Dirac fermions, what requires the addition of three extra SM singlets.)
This possibility has been explored in definite models \cite{ing,tau,tau2}.
Indeed, if we write
\begin{equation}
Y = Y^{(0)} + \varepsilon Y^{(1)} \,,
\end{equation}
with $\varepsilon$ small and $Y^{(0)}$ taking the form
\begin{equation}
Y^{(0)} =
\left( \! \begin{array}{ccc}
y_1 & y_2 & y_3 \\ 
\alpha y_1 & \alpha y_2 & \alpha y_3 \\ 
\beta y_1 & \beta y_2 & \beta y_3  
\end{array} \! \right) \,, 
\end{equation}
where
\begin{equation}
\frac{y_1^2}{M_1} + \frac{y_2^2}{M_2} + \frac{y_3^2}{M_3} =0 \,,
\end{equation}
the term $Y^{(0)} M_R^{-1} Y^{(0)T}$ identically vanishes, and the light
neutrino mass matrix reduces to
\begin{equation}
M_\nu \simeq \varepsilon \left( Y^{(0)} M_R^{-1} Y^{(1)T}
+ Y^{(1)} M_R^{-1} Y^{(0)T} \right) + O(\varepsilon^2) \,.
\end{equation} 
In this case it is also possible to cancel the contributions 
to neutrinoless double $\beta$ decay as well, which otherwise would 
require in general much heavier neutrino masses $m_{N_i} \geq 1$ TeV
\cite{M,nless}. Moreover, this framework can accommodate
leptogenesis by making $\beta \ll \alpha$ \cite{tau,tau2}.
At any rate, constructing models where this structure and the size of $\alpha$,
$\beta$ result from symmetries with a natural breaking
does not seem straightforward.

\section{Heavy neutrino signals at large colliders}
\label{sec:5}

Heavy neutrino signals can conserve or violate lepton number. In the LNC case 
final state angular distributions must be used in order to determine the Dirac
or Majorana nature of the heavy neutrino. On the contrary, LNV signals
certify its Majorana character. (Although lepton number can
be violated by light Majorana neutrino masses, they have no relevance at large
colliders because their effects are suppressed
by powers of  $m_\nu /\sqrt s$.) 
Lepton flavour is also conserved within the SM in the limit of vanishing
neutrino masses. This makes LFV signals interesting
as well. 
SM backgrounds to LNV or LFV processes involve the production of extra
neutrinos in the final state (this is because both lepton number and flavour are
preserved
within the minimal SM, in the limit of vanishing light neutrino masses).
Then, in addition to being higher order processes, they can be
greatly reduced in general requiring the absence of significant missing energy.
Consequently, SM backgrounds to these kind of signals are much smaller than for
the ones conserving lepton number and flavour.
This usually translates in more stringent constraints 
on heavy Majorana neutrinos, and justifies concentrating on LNV processes in
some cases. For each accelerator
it is convenient to classify the possible signals according to their 
LNC and lepton flavour conserving (LFC) character in order to address the
discovery potential  in each case. 

We review the estimates for heavy neutrino production at $e^-p$,
$p \ppbar$, $e^+e^-$, and $e^-\gamma$ colliders in turn, 
restricting ourselves to masses $m_N > M_Z$. Since final state
neutrinos are undetected, the observation of LNV signals requires a change 
of two units in the charge of the leptons involved. This makes $e^-p$ and
$e^- \gamma$, a priori, more adequate to search for Majorana neutrinos, because
the initial state has a single charged lepton. The most significant processes
are $e^- p \to N j$ \cite{buch1,buch2,ing} and 
$e^- \gamma \to N W^-$ \cite{pila}.
For $p \ppbar$ collisions the most interesting process is
$q \bar q' \to \ell^+ N$ (and its charge conjugate).
We extend the
analysis in Refs.~\cite{Han,otherpp1,otherpp2} to Dirac neutrinos,
for which backgrounds
are much larger. Finally, the neutral character
of the initial state in $e^+e^-$ colliders makes this machine equally sensitive
to Dirac and Majorana neutrinos in the process
$e^+ e^- \to N \nu$ \cite{clic}.

It is also important to realise that discovering a heavy neutrino seems to
require its on-shell production.
This is so because its mixing with the SM particles is
rather small, what makes necessary the pole enhancement factor
to observe the heavy neutrino signal over the background.
Additionally, the production of an on-shell heavy neutrino allows to
reconstruct its mass and reduce the backgrounds further.

\subsection{$e^- p$ scattering}

\begin{figure}[htb]
\begin{center}
\mbox{\epsfig{file=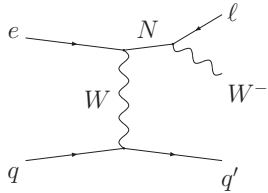,height=2.5cm,clip=}}
\end{center}
\caption{Feynman diagram for the LNV process $e^- q \to \ell^+ W^- q'$. An
additional process $e^- \bar q' \to N \bar q$, obtained interchanging the quark
lines, contributes to $e^- p \to \ell^+ W^- j$ as well.}
\label{fig:ep}
\end{figure}

Heavy neutrinos can be produced in the processes $e^- q \to N q'$,
$e^- \bar q' \to N \bar q$, being $q=u,c$, $q'=d,s$. 
These processes take place with $t$ channel exchange,
and hence their cross section, equal for Dirac and Majorana $N$, is not
suppressed for large $m_N$ by $s$-channel propagators. On the
other hand, $N$ is produced through its mixing with the electron $V_{eN}$,
and heavy neutrinos mixing significantly with the muon or tau but with $V_{eN}
\simeq 0$ are not observable. Depending on the mixing and character of $N$, we
can have the following signals:
\begin{enumerate}
\item For Dirac $N$ coupling only to the electron, the decay $N \to e^- W^+$
gives an SM-like final state $e^- W^+ j$ with a huge background.
\item For Dirac $N$ coupling also with the muon or tau, the decays
$N \to \mu^- / \tau^- \, W^+$ give clean LFV signals
$e^- p \to \mu^- / \tau^- \, W^+ j$.
\item For Majorana $N$, apart from the previous modes we have
$N \to \ell^+ W^-$, yielding a clean LNV signal $e^- p \to \ell^+ W^- j$,
as depicted in Fig.~\ref{fig:ep}.
\end{enumerate}

The $W^\pm$ bosons in the final state can be taken to decay hadronically, what,
in addition to the larger branching ratio, avoids the complication of additional
final state leptons which may hide the non-conservation of lepton number
and/or flavour.
The LNV signal $e^- p \to e^+ jjj$ of a heavy Majorana neutrino has been
studied in the literature \cite{buch1,buch2,ing}. 
At the centre of mass (CM) energy $\sqrt s = 314$ GeV available at HERA
heavy neutrino
production cross sections are very small, due to the suppression of the parton
density functions (PDFs) at high $x$. We have rescaled the limits obtained in
Ref.~\cite{ing}, saturating the improved upper bound $|V_{eN}|^2 \leq 0.0054$.
For a luminosity of 200 pb$^{-1}$,
assuming no backgrounds and a perfect detection efficiency, HERA could give $2
\sigma$ evidence (conventionally taken as 3 signal events in the absence of
background) for a 100 GeV
neutrino coupling only to the electron. This sensitivity is similar to the one
achieved at LEP \cite{l3}.

A hypothetical $\mathrm{LEP} \otimes \mathrm{LHC}$ $ep$ machine, with
$\sqrt s = 1.3$ TeV and a
luminosity of 2 fb$^{-1}$ per year, could extend the LHC sensitivity for heavy
Majorana neutrinos. Taking as statistical criterion for
$5 \sigma$ discovery the observation of 10 signal events, this significance
would be achieved for heavy neutrino masses up to $\sim 550$ GeV, doubling the
LHC reach (see next subsection).
If no signal is found, the bounds $|V_{eN}| \lesssim 0.02$ at 90\% CL could be
set for
$m_N \simeq 300$ GeV, and useful constraints would be obtained up to $m_N \sim
700$ GeV.

\subsection{$p \ppbar$ collisions}

\begin{figure}[htb]
\begin{center}
\begin{tabular}{ccc}
\mbox{\epsfig{file=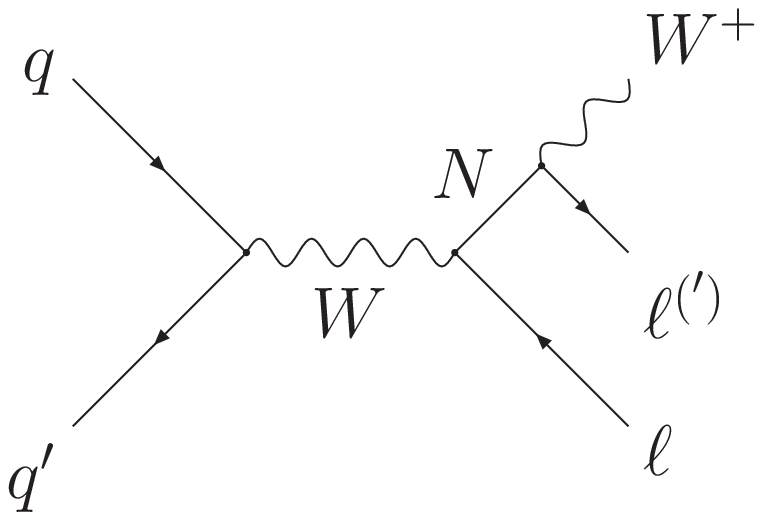,height=2.5cm,clip=}} & \quad \quad &
\mbox{\epsfig{file=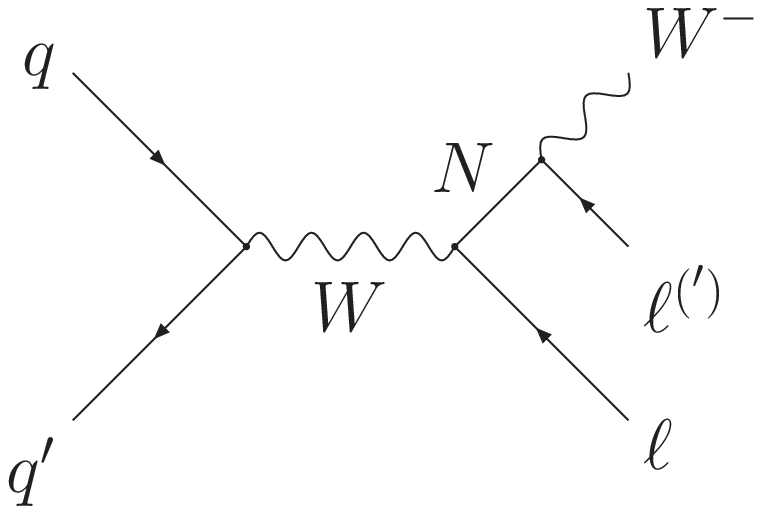,height=2.5cm,clip=}} \\[1mm]
(a) & & (b)
\end{tabular}
\end{center}
\caption{Feynman diagrams for the process $q \bar q' \to \ell^+ N$, followed by
LNC decay $N \to \ell^{(')-} W^+$ (a) and LNV decay $N \to \ell^{(')+ }W^-$ (b).
Additional diagrams with off-shell $N$ contributing to the same final state are
not shown. For the charge conjugate processes the diagrams are analogous.}
\label{fig:pp1}
\end{figure}

Hadron colliders can produce heavy neutrinos in association with a charged
lepton in $q \bar q' \to \ell^+ N$, as shown in Fig.~\ref{fig:pp1}
(plus its charge conjugate $\bar q q' \to \ell^- N$).
This process is relevant for moderate $N$ masses. It is
mediated by $s$-channel $W$ exchange, and then
for large $m_N$ it is suppressed not only by PDFs but also by the
$W$ propagator
(in contrast to $ep \to Nj$ scattering previously discussed). However,
one important advantage compared to other colliders is that $\ell^+ N$
production at LHC
does not require a sizeable coupling to the electron, as it can be observed from
Fig.~\ref{fig:pp1}. It can lead to the
following signals:
\begin{enumerate}
\item For a Dirac $N$ mixing with only one lepton flavour, the decay
$N \to \ell^- W^+$ yields a $\ell^+ \ell^- W^+$ final state, with a huge
SM background.

\item For a Dirac $N$ coupled to more than one charged lepton we can also have
$N \to \ell^{'-} W^+$ with $\ell' \neq \ell$, giving the LFV signal
$\ell^+ \ell^{'-} W^+$ shown in Fig.~\ref{fig:pp1} (a), which has much
smaller backgrounds.

\item For a Majorana $N$, in addition to LNC signals we have LNV ones arising
from the decay $N \to \ell^{(')+} W^-$ in Fig.~\ref{fig:pp1} (b), which have
small backgrounds too.
\end{enumerate}

The charge conjugate process $\bar q q' \to \ell^- N$ yields the charge
conjugate final states. For the same values of the couplings,
cross sections for on-shell $N$ production are the same for heavy Dirac and
Majorana neutrinos (for the latter, additional diagrams with off-shell $N$ can
mediate
the final states considered, but their contribution is very small). Therefore,
since a Majorana $N$ has LNV decays as well as LNC ones, the relation
\begin{equation}
\sigma_\mathrm{D} \, (p \ppbar \to \ell^\pm \ell^{(') \mp} W^\pm) \simeq
2 \sigma_\mathrm{M} \, (p \ppbar \to \ell^\pm \ell^{(')\mp} W^\pm) \simeq 
2 \sigma_\mathrm{M} \, (p \ppbar \to \ell^\pm \ell^{(')\pm} W^\mp) 
\end{equation}
holds, although
$\sigma_\mathrm{M} \, (pp \to \ell^+ \ell^{(')+} W^-)
\neq \sigma_\mathrm{M} \, (pp \to \ell^- \ell^{(')-} W^+)$ in $pp$
collisions due to the different PDFs involved.

Apart from angular distributions, the obvious difference between the
processes (i)--(iii) above is the SM background, huge for the first
case and small for the other two.
In order to estimate the LHC discovery potential
we have implemented in ALPGEN \cite{ALPGEN} the relevant vertices presented in
section \ref{Introduction}, finding results consistent with previous 
analyses for Majorana neutrinos \cite{Han}. 
We will present a detailed study of signals and backgrounds 
with a proper simulation of the experimental detection elsewhere.
For our numerical estimates we assume a heavy neutrino $N$ coupling only
to muons and saturating the present limit $|V_{\mu N}|^2 \leq 0.0096$,
which yields $\mu^\pm \mu^\mp W^\pm$ final states and, provided $N$ is a
Majorana fermion, $\mu^\pm \mu^\pm W^\mp$ as well.
This is the most interesting situation for LHC, where it can outperform
other planned $e^+ e^-$ and $e^- \gamma$ colliders, as it will be shown later.
The analysis if $N$ couples
only to the electron is completely analogous, while if it couples to the tau
the decays and tagging efficiency of this lepton must be considered in the
analysis.
We select $W^\pm$ hadronic decays, and fix the Higgs mass $M_H = 120$ GeV.
The processes considered are 
\begin{eqnarray}
p \ppbar & \to & \mu^+ \mu^- j j \,, \quad \quad {\rm (LNC)} \nonumber \\
p \ppbar & \to & \mu^\pm \mu^\pm j j  \,. \quad \quad {\rm (LNV)}
\label{eq:proc1}
\end{eqnarray}
Among the most relevant SM backgrounds we select as example
\begin{equation}
p \ppbar \to  Z/\gamma^* jj \to \mu^+ \mu^- jj
\end{equation}
in the LNC case, and
\begin{eqnarray}
p \ppbar \to W^\pm W^\pm W^\mp \to \mu^\pm \mu^\pm jj~+~{\rm missing~energy}
\end{eqnarray}
in the LNV one, computing them with ALPGEN. 
In the former case (LNC) the background is huge. For instance,
the signal cross section without cuts for
$p \ppbar \rightarrow \mu^\pm N \rightarrow \mu^\pm \mu^\mp W^\pm 
\rightarrow \mu^\pm \mu^\mp jj$ at LHC is 86 fb for $m_N= 100$ GeV, 
while the $Zjj$ background is around 400 times larger.
The cuts adopted to simulate the detector coverage and particle / jet isolation
are
\begin{eqnarray}
& p_t^\mu > 20 \; {\rm GeV} \,, & p_t^j > 30 \; {\rm GeV} \,, \nonumber \\
& |\eta _{\mu,j}| < 2.5 \,, & \Delta R_{jj, j\mu, \mu\mu} > 0.4 \,,
\label{cuts}
\end{eqnarray}
with $p_t$ the transverse momentum, $\eta$ the pseudorapidity and $\Delta R$ the
lego-plot distance, in standard notation. Besides these minimal
``pre-selection'' criteria we impose cuts on various final state invariant
masses $M$ to suppress backgrounds. In the LNC case we require
\begin{eqnarray}
& 60 \; {\rm GeV} < M_{jj} < 100 \; {\rm GeV} \,,  \nonumber \\
&  40 \; {\rm GeV} < M_{\mu\mu} < 70 \; {\rm GeV}  \;\;\;\; {\rm or} \;\;\;\; 
 M_{\mu\mu} > 110 \; {\rm GeV} \,,
\label{cutb}
\end{eqnarray}
in order to reconstruct the $W$ boson from $N$ decay and reduce the
$Z / \gamma^*$ contributions. In the LNV case we ask
\begin{eqnarray}
& 60 \; {\rm GeV} < M_{jj} < 100 \; {\rm GeV} \,,  \nonumber \\
& \ptmiss < 20 \; {\rm GeV} \,, 
\label{cutbb}
\end{eqnarray}
so as to reconstruct the $W$ boson and take advantage  of the fact that the
signal has no significant missing momentum $\ptmiss$. Moreover, as the
observation of a
heavy neutrino requires its mass reconstruction, we also impose in both cases
\begin{eqnarray}
 0.9 \; m_N < M_{jj\mu} < 1.1 \; m_N \,,  
\label{cutr}
\end{eqnarray}
for at least one of the two $\mu$ assignments. The corresponding cross sections
are given in Tables \ref{tab:lhc} and \ref{tab:tev} for LHC and Tevatron,
respectively. For the LNC signal the heavy neutrino is assumed to have Dirac
nature
(for a Majorana neutrino this cross section would be roughly one half).
We have concentrated on the mass region $m_N > M_Z$ where the
signal cross section gets smaller (the case $m_N < m_Z$ has been also
considered for  $p \ppbar \to \, \mu^\pm \mu^\pm j j$ in Ref. \cite{Han}).
As we stressed at the beginning of this subsection, cross sections are
suppressed for larger $m_N$ values. We observe for example that changing
$m_N= 100$ GeV to $m_N= 500$ GeV implies a signal reduction of almost 2 orders
of magnitude at LHC. 

\begin{table}[htb]
\begin{center}
\begin{tabular}{lcccc}
$m_N$ & $\begin{array}{c} \mu^+ \mu^- jj \\ {\rm signal} \end{array}$ 
& $\begin{array}{c} Zjj \\ {\rm background} \end{array}$ 
& $\begin{array}{c} \mu^\pm \mu^\pm jj \\ {\rm signal} \end{array}$
& $\begin{array}{c} W^\pm W^\pm W^\mp \\ {\rm background} \end{array}$ \\
\hline
100  & 2.6  & 43  & 2.0  & 0.0012 \\
200  & 0.83  & 91  & 0.48  & 0.0044 \\
300  & 0.29  & 44  & 0.16  & 0.0023 \\
400  & 0.13  & 22  & 0.068 & 0.0012 \\
500  & 0.066 & 11  & 0.034 & 0.0007
\end{tabular}
\caption{Signal and background cross sections (in fb) 
as a function of the heavy neutrino mass (in GeV) at LHC ($\sqrt{s}= 14$ TeV) 
for the cuts given in the text.}
\label{tab:lhc} 
\end{center}
\end{table}
\begin{table}[htb]
\begin{center}
\begin{tabular}{lcccc}
$m_N$ & $\begin{array}{c} \mu^+ \mu^- jj \\ {\rm signal} \end{array}$ 
& $\begin{array}{c} Zjj \\ {\rm background} \end{array}$ 
& $\begin{array}{c} \mu^\pm \mu^\pm jj \\ {\rm signal} \end{array}$
& $\begin{array}{c} W^\pm W^\pm W^\mp \\ {\rm background} \end{array}$ \\
\hline
100  & 0.51   & 2.9 & 0.40   & 0.0001 \\
200  & 0.12   & 4.9 & 0.071  & 0.0004 \\
300  & 0.025  & 1.7 & 0.014  & 0.0001 \\
400  & 0.0060 & 0.68 & 0.0032 & 0.00005 \\
500  & 0.0015 & 0.24 & 0.0008 & 0.00001 \\
\end{tabular}
\caption{ The same as in Table \ref{tab:lhc} but at Tevatron 
($\sqrt{s}= 1.96$ TeV).}   
\label{tab:tev}
\end{center}
\end{table} 

Tevatron does not seem to have a chance to observe a new heavy 
neutrino in this mass range with a luminosity of 2 fb$^{-1}$. For LHC,
although the comparison between the first two columns of Table~\ref{tab:lhc}
seems to indicate that it may be difficult to observe a Dirac heavy neutrino,
one could still carefully refine the cut selection to
improve the signal to background significance. In this way neutrino masses
$\sim 100$ GeV might be observable. 
The LNV process clearly offers better prospects to detect a heavy Majorana 
neutrino. In this case one expects to discover with $5 \sigma$ significance (10
signal events without background) a Majorana neutrino coupling only to the muon
with a mass up to $350$ GeV in this channel, 
assuming a luminosity of 100 fb$^{-1}$. This result is in agreement with
previous ones \cite{Han}.
Conversely, if no signal is observed present bounds on the mixing angles are
improved. For instance, for $m_N = 200$ GeV we would obtain the 90\% CL upper
limit
$|V_{\mu N}| \leq 0.022$, improving the present bound by a factor of four.
(This limit is obtained assuming no observed events, what yields an upper
limit of 2.44 for the signal in the absence of background \cite{fc}.)
Obviously, further backgrounds must be taken into account and a realistic
detector simulation performed, especially regarding the background suppression
with the requirement $\ptmiss < 20$ GeV, which seems more delicate.
On the other hand, the kinematical cuts used to enhance the signal significance
can still be improved.

For a heavy Majorana neutrino simultaneously coupling to the electron and muon 
the results (summing all relevant signals) are expected to be similar, since
the backgrounds involving
electrons and muons have the same size. In case that $N$ couples significantly
to the tau lepton, results will be worse because decays involving taus are
more difficult to tag.
An interesting possibility is that of a heavy Dirac neutrino coupling to more
than one charged lepton. In this case, LFV signals as described in point (ii)
above have much smaller SM backgrounds than LFC ones discussed here, and 
better limits could be obtained.

\subsection{$e^+ e^-$ annihilation}

\begin{figure}[htb]
\begin{center}
\begin{tabular}{ccccc}
\mbox{\epsfig{file=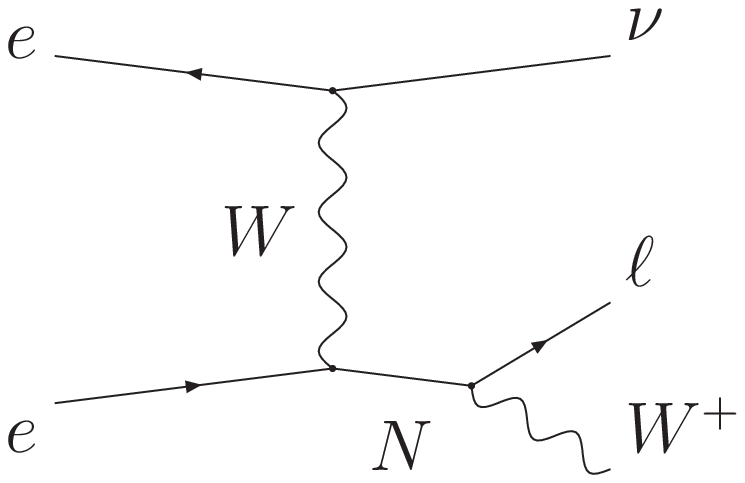,height=2.5cm,clip=}} & \quad &
\mbox{\epsfig{file=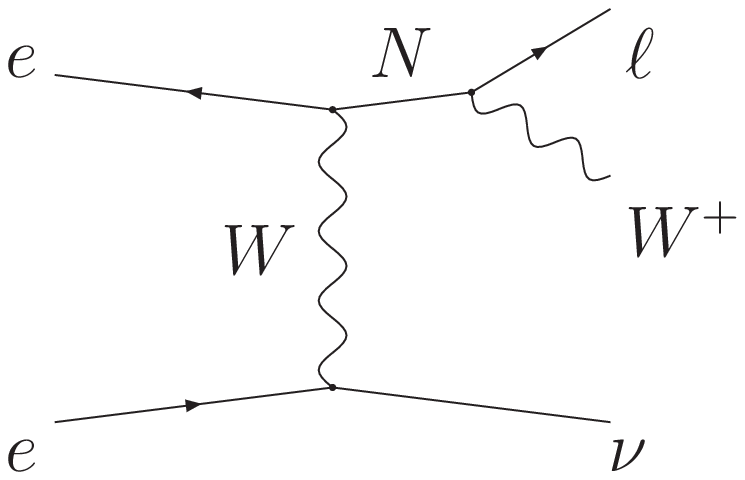,height=2.5cm,clip=}} & \quad &
\mbox{\epsfig{file=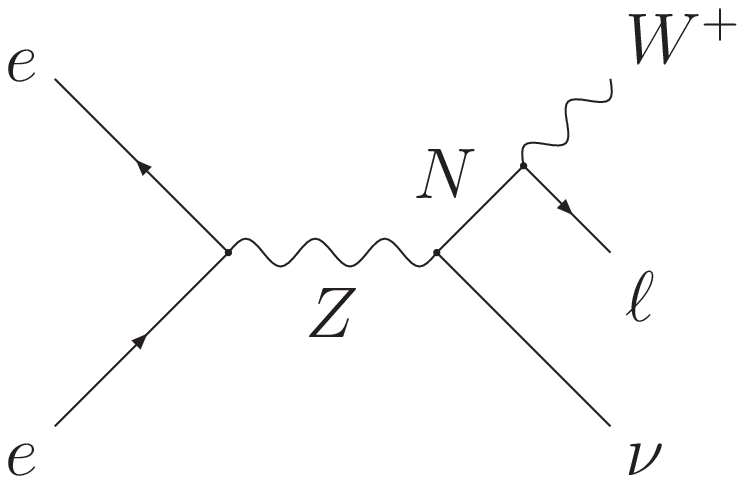,height=2.5cm,clip=}} \\
(a) & & (b) & & (c)
\end{tabular}
\end{center}
\caption{Feynman diagrams for the process $e^+ e^- \to \ell^- W^+ \nu$
involving on-shell Majorana $N$ exchange. Diagram (b) is only present if $N$ is
a Majorana fermion.
Additional diagrams with off-shell $N$ are not shown.}
\label{fig:ee1}
\end{figure}

The process $e^+ e^- \to N \nu$ can produce heavy neutrinos which couple to
the electron, for masses up to nearly the kinematical limit imposed by
CM energy \cite{zerwas,GZ2,djouadi,GZ4}.
In the Dirac case both the neutrino $N$ and antineutrino $\bar N$
are produced (with equal cross sections), and a Majorana $N$ is produced with a
cross section two times larger. The subsequent decays $N \to \ell^- W^+$,
$\bar N \to \ell^+ W^-$ for a Dirac neutrino and $N \to \ell^- W^+$,
$N \to \ell^+ W^-$ for a Majorana one yield $\ell^\pm W^\mp$ final states,
with total cross
sections $\sigma (e^+ e^- \to \ell^- W^+ \nu) = 
\sigma (e^+ e^- \to \ell^+ W^- \nu)$. Additionally, these two cross sections are
almost independent of the Dirac or Majorana nature of the produced neutrino.
The diagrams for $e^+ e^- \to \ell^- W^+ \nu$ mediated by on-shell $N$ exchange
are shown in Fig.~\ref{fig:ee1}.\footnote{If $N$ does not couple to the electron
it can still be produced through the diagram (c), but this contribution is very
suppressed by the $s$-channel $Z$ propagator, and does not lead to an observable
signal \cite{plb}.} 
The SM background is given by
$e^+ e^- \to \ell^- W^+ \nu$, including resonant $W^+ W^-$ production and 
several other Feynman diagrams, which dominate at high energies. We observe that
the only indication of lepton number violation
in diagram \ref{fig:ee1} (b) is the helicity of the final state neutrino, which
remains undetected. Therefore, there is no advantage for LNV processes in what
respects to background reduction, and limits obtained are the same for Dirac and
Majorana fermions.

\begin{figure}[htb]
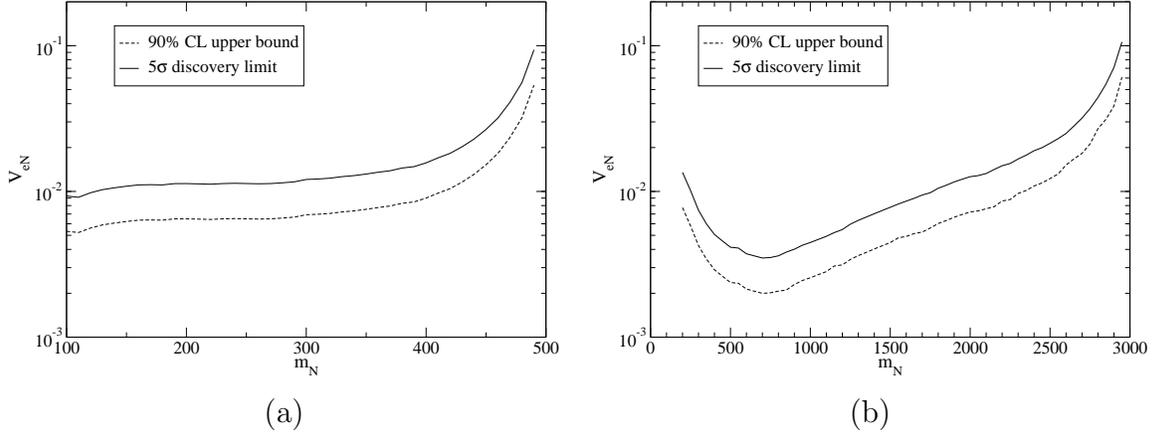

\begin{center}
\begin{tabular}{cc}
\epsfig{file=Figs/mass-coup-ILC.eps,height=5cm,clip=} & 
\epsfig{file=Figs/mass-coup.eps,height=5cm,clip=} \\
(a) & (b)
\end{tabular}
\caption{Dependence of the discovery and upper limits on
$V_{eN}$ on the heavy neutrino mass, for ILC (a) and CLIC (b). Both plots
assume mixing only with the electron.}
\label{fig:mass-coup}
\end{center}
\end{figure}

We consider $e^+ e^-$ annihilation at a CM energy of 500 GeV, as proposed for an
international linear collider (ILC) and 3 TeV, as might be the case of a future
compact linear collider (CLIC), examining their discovery potential for heavy
neutrinos \cite{clic}. Since they are produced through its mixing with the
electron, two cases are worth discussing for this process: (a) $N$ only couples
to the
electron; (b) $N$ also mixes with the muon or tau lepton. $W$ hadronic decays
are selected, and the luminosities for ILC and CLIC are of 345 fb$^{-1}$ and
1000 fb$^{-1}$, respectively, corresponding to one year of running. Beam
polarisations $P_{e^+} = 0.6$, $P_{e^-} =
-0.8$ are also used. The discovery potential for both machines is
summarised in Fig.~\ref{fig:mass-coup}. (Notice that the mass range in the
second plot has been enlarged with respect to Ref.~\cite{clic}.)

For ILC the sensitivity is nearly the same for masses between 100 and 400 GeV,
because the reduction of the background at larger transverse
momenta makes up for the
decrease in signal cross sections for larger $m_N$. A heavy neutrino with a
coupling $|V_{eN}| \geq 0.01$ could be discovered with $5 \sigma$ significance,
and if no signal is seen the limit $|V_{eN}| \leq 0.006$ could be set
at 90\% CL, improving
the eventual LHC bounds by a factor of 4. CLIC would be able to explore $N$
masses in the TeV range, as it can be observed in Fig.~\ref{fig:mass-coup} (b),
and provide very stringent limits $|V_{eN}| \leq 0.001-0.002$ for
$m_N = 400-1000$ GeV.
These estimates have been obtained at the partonic level and with a simple
simulation of the detector resolution. Therefore, it is expected that with a
detailed simulation the results will be a little worse.
We also remark that the Dirac or Majorana nature of a 
heavy neutrino eventually discovered could be unambiguously established with
the analysis of its opening angle distribution \cite{clic}.

\begin{figure}[htb]
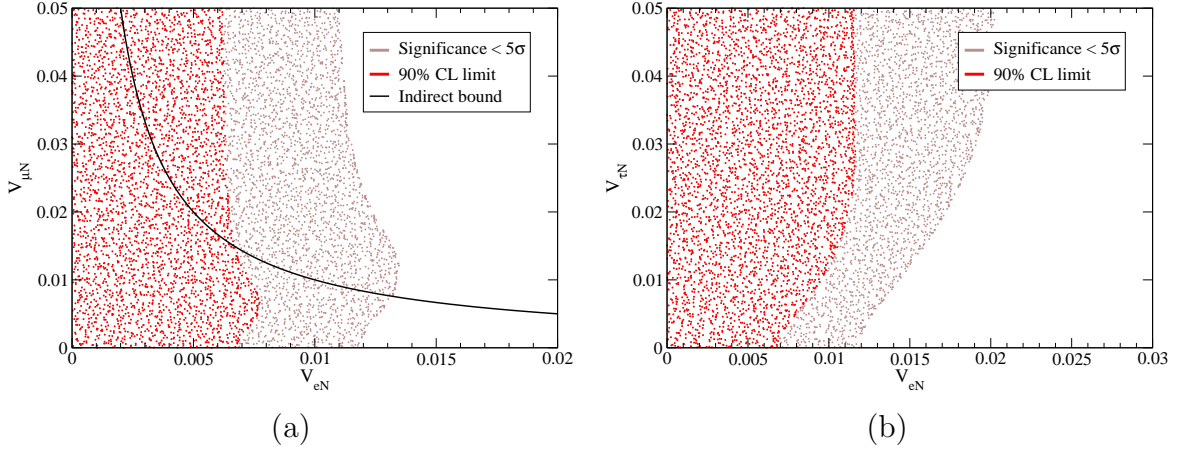

\begin{center}
\begin{tabular}{cc}
\epsfig{file=Figs/bound-em-ILC.eps,height=5.2cm,clip=} &
\epsfig{file=Figs/bound-et-ILC.eps,height=5.2cm,clip=} \\
(a) & (b)
\end{tabular}
\caption{Combined limits obtained at ILC on: $V_{eN}$ and $V_{\mu N}$, for
$V_{\tau N} = 0$ (a); $V_{eN}$ and $V_{\tau N}$, for $V_{\mu N} = 0$ (b). The
red areas represent the 90\% CL limits if no signal is observed. The white areas
extend up to present bounds $V_{eN} \leq 0.073$, $V_{\mu N} \leq 0.098$,
$V_{\tau N} \leq 0.13$, and correspond to the region where a combined
statistical significance of $5\sigma$ or larger is achieved. The indirect limit
from $\mu - e$ LFV processes is also shown. We take $m_N = 300$ GeV.}
\label{fig:limits-ILC}
\end{center}
\end{figure}

For heavy neutrinos mixing also with the muon or the tau, additional signals
$\mu^\pm W^\mp \nu$ or $\tau^\pm W^\mp \nu$ are present.
The discovery potential for ILC in this case is shown in
Fig.~\ref{fig:limits-ILC}, considering mixing either with the muon (a)
or with the $\tau$ (b). In the first case the combined limits do not depend on
the mixing with the muon, essentially due to the fact that backgrounds involving
electrons and muons have the same size. In the second case the limits are worse
for larger $|V_{\tau N}|$ because in this situation the heavy neutrino has a
larger branching fraction to $\tau^\pm W^\mp$,
and $\tau$ detection is experimentally more difficult.
We have also included the constraints from low-energy LFV processes for
comparison, which are relevant only for $N$ coupling simultaneously
to $e$ and $\mu$.
The direct limit from $N$ production complements the indirect ones
for small $|V_{\mu N}|$.

\begin{figure}[htb]
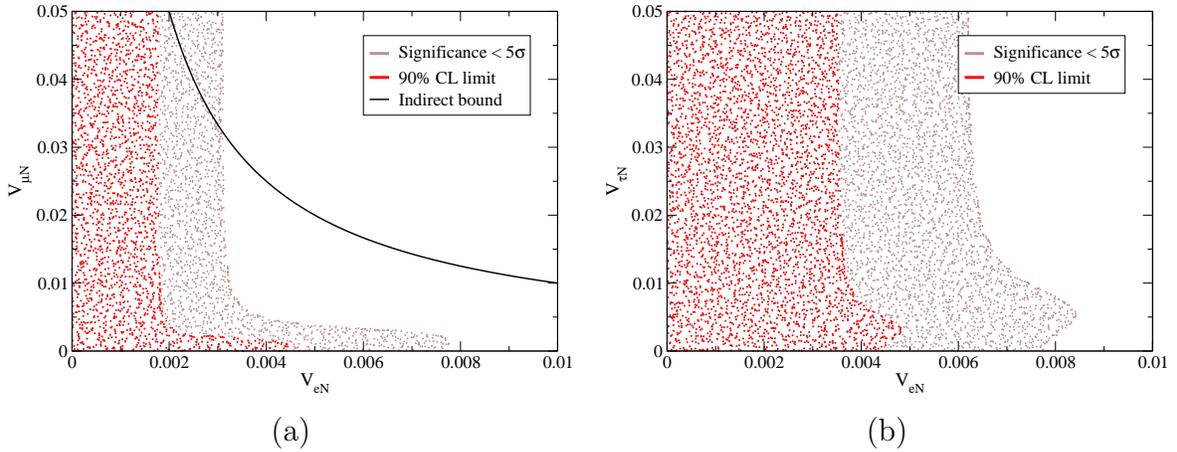

\begin{center}
\begin{tabular}{cc}
\epsfig{file=Figs/bound-em.eps,height=5.2cm,clip=} &
\epsfig{file=Figs/bound-et.eps,height=5.2cm,clip=} \\
(a) & (b)
\end{tabular}
\caption{The same as in Fig.~\ref{fig:limits-ILC} but for CLIC,
taking $m_N = 1500$ GeV.}
\label{fig:limits}
\end{center}
\end{figure}

At CLIC energies the behaviour is completely different. Since backgrounds
involving muons or taus are much smaller \cite{clic}, mixing with a second
charged lepton increases the observability of the heavy neutrino.
The combined limits are shown in Fig.~\ref{fig:limits}, for mixing with the muon
(a) or the tau (b). In the first case the limits are greatly improved for
$|V_{\mu N}| \gtrsim 0.005$, while in the second case the effect is partially
compensated
by the worse detection of $\tau$ leptons. We also point out that the direct
limit eventually obtained from $N$ production is far better than the indirect
one from present
low-energy LFV processes, and would remain competitive with future
improvements of the upper bounds on $\mathrm{Br}(\mu \to e \gamma)$ \cite{MEG}
and $\mu - e$ conversion in nuclei \cite{MECO}.

\begin{figure}[htb]
\begin{center}
\begin{tabular}{ccccc}
\mbox{\epsfig{file=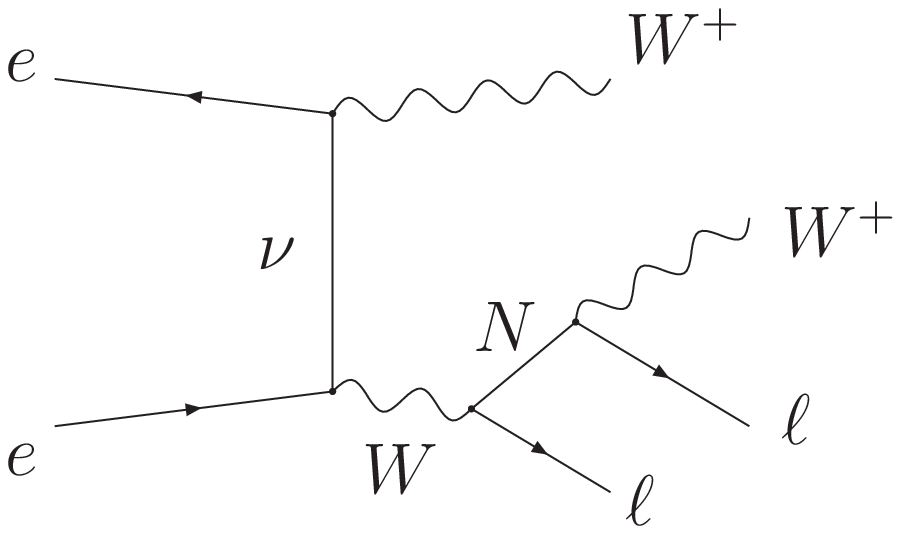,height=2.5cm,clip=}} & \quad &
\mbox{\epsfig{file=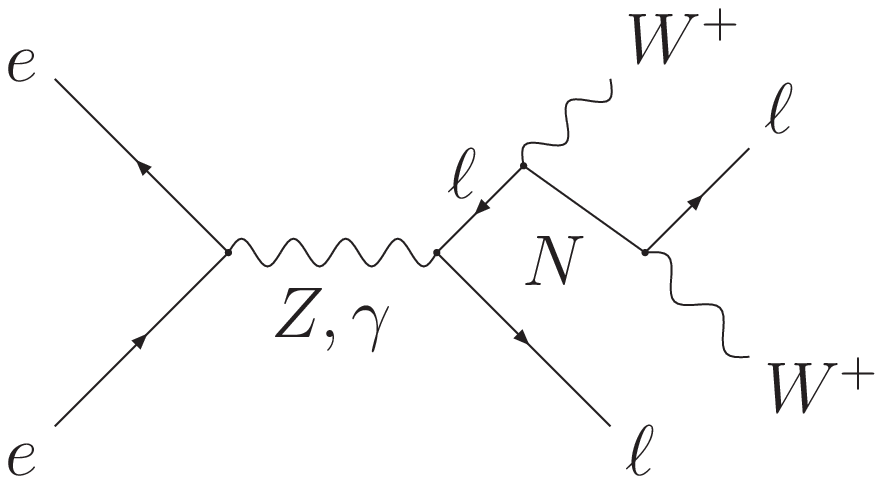,height=2.5cm,clip=}} & \quad &
\mbox{\epsfig{file=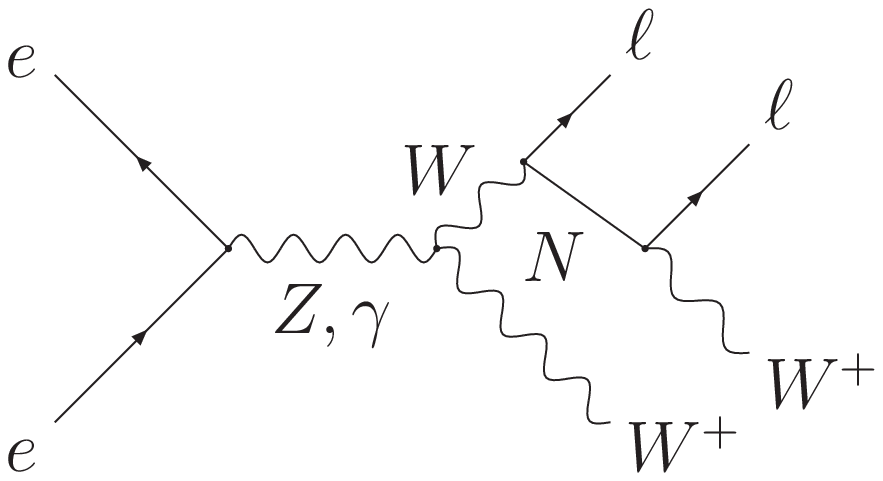,height=2.5cm,clip=}}
\end{tabular}
\end{center}
\caption{Feynman diagrams for the process $e^+ e^- \to \ell^- \ell^- W^+ W^+$
involving on-shell Majorana $N$ exchange.
Additional diagrams with off-shell $N$ are not shown.}
\label{fig:ee2}
\end{figure}

A heavy neutrino $N$ with $V_{eN} \simeq 0$ can still be produced in $e^+ e^-$
annihilation through
the higher-order process $e^+ e^- \to \ell^- N W^+$ (and its charge conjugate).
Its cross section is much smaller than
for $e^+ e^- \to N \nu$, however. Depending on the nature and mixing of $N$,
we can have the following final states:
\begin{enumerate}
\item For Dirac $N$ coupling only to the muon or tau, $N \to \ell^+ W^-$
($\ell=\mu,\tau$) gives an SM-like $\ell^+ \ell^- W^+ W^-$ signal 
with a large background.
\item A Dirac $N$ coupling to $\mu$ and $\tau$ also yields the LFV
final state $\mu^\pm \tau^\mp W^+ W^-$, for which the background is much
smaller.
\item For a Majorana $N$, the decay $N \to \ell^- W^+$ leads to a LNV signal
$\ell^- \ell^- W^+ W^+$, as shown in Fig.~\ref{fig:ee2}. SM background is very
small as well in this case.
\end{enumerate}

We have estimated the ILC discovery potential in the latter case (iii),
for $m_N = 200$ GeV, $V_{eN} = V_{\tau N} = 0$, $|V_{\mu N}|^2 = 0.0096$.
Assuming perfect
detection efficiency, 8 $\mu^\pm \mu^\pm jjjj$ events could be obtained within
one year of running. The SM background is given by $e^+ e^- \to W^- W^- W^+ W^-
\to \mu^\pm \mu^\pm \nu \nu jjjj$, and
it is assumed that it can be reduced to negligible levels 
(what must be confirmed with a detailed simulation)
requiring the absence
of significant missing momentum and the reconstruction of the $N$ invariant
mass. Hence, the heavy neutrino could be clearly observed but with a
significance smaller than $5 \sigma$. If no signal is found, the limit 
$|V_{\mu N}| \leq 0.055$ could be set at 90\% CL, a factor of two lower than the
one obtained at LHC.

\subsection{$e^- \gamma$ collisions}

\begin{figure}[htb]
\begin{center}
\begin{tabular}{ccc}
\mbox{\epsfig{file=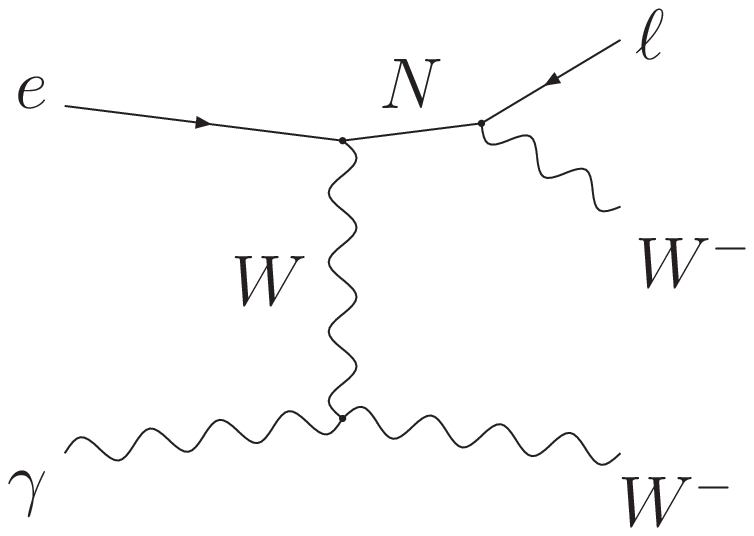,height=2.5cm,clip=}} & \quad \quad &
\mbox{\epsfig{file=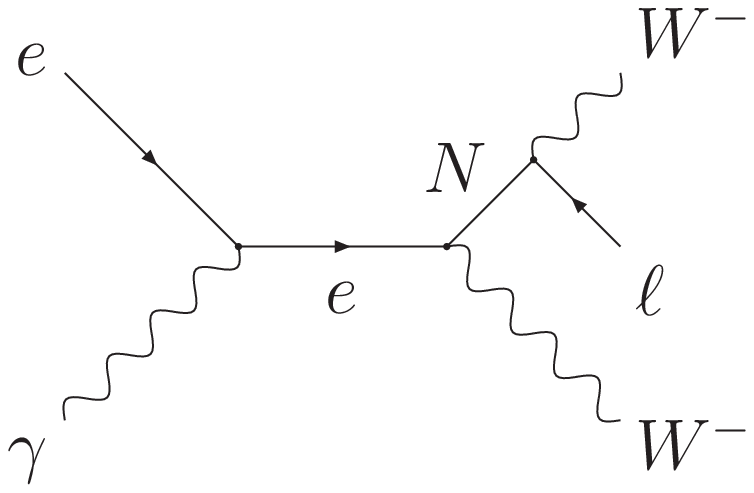,height=2.5cm,clip=}}
\end{tabular}
\end{center}
\caption{Feynman diagrams for the LNV process $e^- \gamma \to \ell^+ W^- W^-$.
Additional diagrams with off-shell $N$ are not shown.}
\label{fig:eg1}
\end{figure}

A proposed option for ILC is to have $e^- \gamma$ collisions at CM energies
of several hundreds of GeV.
From the point of view of heavy neutrino physics, this would be a very
interesting
possibility, complementing the capabilities of $e^+ e^-$
annihilation. Heavy neutrinos can be produced in the process $e^- \gamma \to N
W^-$. The cross section is the same for Dirac and Majorana $N$, but 
depending on its character and mixing with the charged leptons we
can have the following final states:
\begin{enumerate}
\item For a Dirac $N$ coupling only to the electron, $N \to e^- W^+$ gives a
SM-like signal $e^- \gamma \to e^- W^+ W^-$ with a large background.
\item For a Dirac $N$ coupling also to the muon or tau, we can have 
$N \to \mu^- / \tau^- W^+$. The resulting LFV signals $e^- \gamma \to 
\mu^- / \tau^- W^+ W^-$ are not present in the SM, hence their backgrounds are
small.  
\item For a Majorana $N$, apart from these final states we have
$N \to \ell^+ W^-$, giving a clean LNV signal $e^- \gamma \to \ell^+ W^- W^-$,
shown in Fig.~\ref{fig:eg1}.
\end{enumerate}

The $W$ bosons can be taken to decay hadronically, what in addition to the
larger branching ratio avoids the complication of additional final state
leptons. SM backgrounds to the LFV and LNV signals involve additional final
state neutrinos, and they can be reduced by requiring the absence of
significant missing
transverse momentum. This is however a delicate issue and requires detailed
simulations to confirm the parton-level expectations.

The process (iii), for a Majorana $N$ coupling mainly with the electron,
has been studied in Ref.~\cite{pila}. The SM background is given by
$e^- \gamma \to W^+ W^- W^- \nu_e \to e^+ \nu_e W^- W^- \nu_e$.
For a CM energy of 500 GeV and a luminosity of 100 fb$^{-1}$,
$5 \sigma$ evidence (taking as criterion the production of 10 signal
events) for a 200 GeV Majorana neutrino could be achieved
for mixings $|V_{eN}| \geq 0.0046$. Conversely, if no signal is observed the
bound $|V_{eN}| \leq 2.3 \times 10^{-3}$ can be set at 90\% CL.
These limits assume that SM background can be
essentially eliminated without affecting the signals, what may be too
optimistic. Thus $e^- \gamma$ collisions improve
the ILC limit by a factor of two for $m_N$ around this value. For heavier $N$
the production cross section decreases quickly, and for $m_N =
400$ GeV the limits for $e^+ e^-$ and $e^- \gamma$ collisions are similar.
On the other hand, limits for heavy Dirac neutrinos are much better at ILC.

\begin{figure}[htb]
\begin{center}
\begin{tabular}{ccccc}
\mbox{\epsfig{file=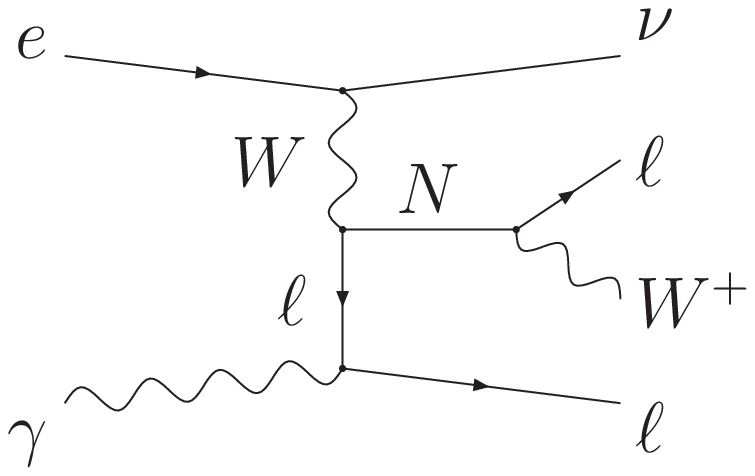,height=2.5cm,clip=}} & \quad &
\mbox{\epsfig{file=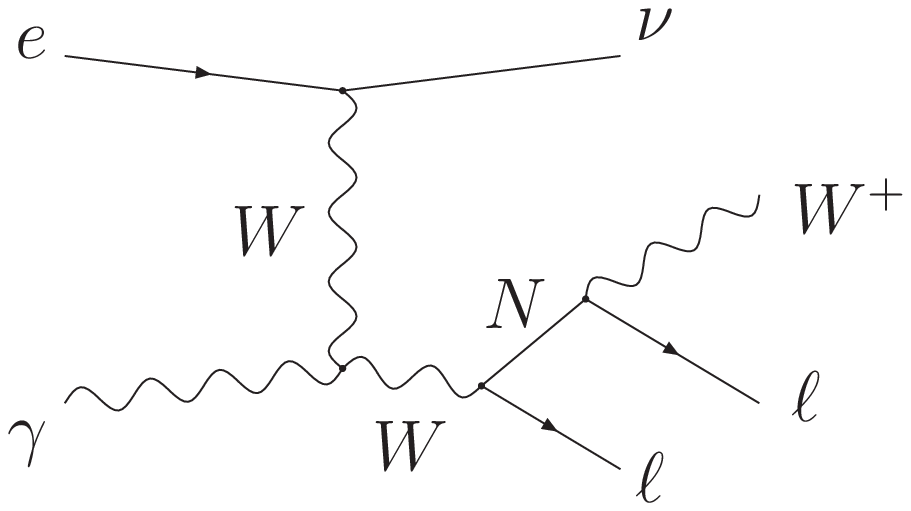,height=2.5cm,clip=}} & \quad &
\mbox{\epsfig{file=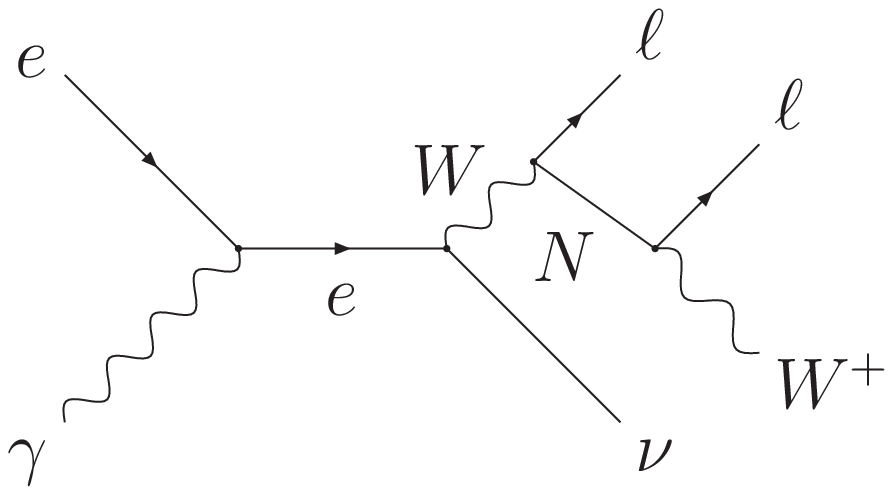,height=2.5cm,clip=}}
\end{tabular}
\end{center}
\caption{Feynman diagrams for the process
$e^- \gamma \to \ell^- \ell^- W^+ \nu$ involving on-shell Majorana $N$ exchange.
Additional diagrams with off-shell $N$ are not shown.}
\label{fig:eg2}
\end{figure}

As in $e^+ e^-$ annihilation, heavy neutrinos which do not couple
to the electron can be produced in a sub-leading process, in this case
$e^- \gamma \to N \ell^- \nu$, $\ell = \mu,\tau$. The following final states
are possible:
\begin{enumerate}
\item For Dirac $N$ coupling only to $\ell = \mu$ or $\ell = \tau$,
$N \to \ell^+ W^-$ gives a signal $\ell^+ \ell^- W^- \nu$ which has a large SM
 background
\item A Dirac $N$ coupling to both can give a LFV signal $\mu^\pm \tau^\mp W^-
\nu$ which is easier to detect
\item For a Majorana $N$, the decay $N \to \ell^- W^+$ leads to a LNV signal
$\ell^- \ell^- W^+ \nu$, as shown in Fig.~\ref{fig:eg2}. SM background is very
small in this case.
\end{enumerate}

The third case has been studied in Ref.~\cite{pila}. Assuming that the
background can be essentially eliminated with kinematical cuts (a fact which
must be confirmed with a detailed simulation), a 200 GeV neutrino coupling only
to the muon can be discovered at $5 \sigma$ level for $|V_{\mu N}| \geq 0.09$.
If no signal is observed, the bound $|V_{\mu N}| \leq 0.045$ can be set. These
figures improve slightly the ones from $e^+ e^-$ annihilation at 500
GeV, shown in the previous subsection, but are still a factor of two
worse than the ones achievable at LHC.

\section{Conclusions}
\label{sec:6}

At present, neutrinos are known to be massive. The minimal SM extension
necessary to account for this experimental fact includes three
ultra-heavy eigenstates with masses of the order of $10^{13}$ GeV, which 
lead to light neutrino masses via the seesaw mechanism.
These states are directly unobservable. Moreover, at low energies the 
number of parameters in the neutrino sector (when heavy states are integrated
out) is smaller than at the high scale, and this minimal seesaw mechanism
is untestable (for a discussion see Ref.~\cite{sd}).
This has given extra motivation for models lowering to the TeV the scale of new
physics which originates light neutrino masses. (Other possibility to obtain
predictive models is to introduce extra symmetries.)
The new heavy neutrinos appearing at this scale could be
directly observed in future colliders, provided their mixing with the charged
leptons is $O(10^{-2})$ or larger.
With a completely different motivation in mind, extra-dimensional and Little
Higgs models are also proposed, in the first case aiming to reduce the huge
hierarchy
between the electroweak and Planck scales, and in the second to cancel large
corrections to the Higgs boson mass.
Both classes of models can have new heavy neutrinos in their additional
particle spectrum, light enough to be observable.

Independently of their origin, the mixing of heavy neutrinos is constrained by
low-energy data, including lepton universality, LFV processes and
neutrinoless double $\beta$ decay. Additionally, their seesaw-like contributions to
light neutrino masses must be kept under control.
Model-independent bounds from universality
restrict their mixing with a given charged lepton, while LFV
processes constrain the simultaneous mixing to more
than one charged lepton. Neutrinoless double $\beta$ decay imposes a strong
constraint on $V_{e N}$ if only one heavy Majorana neutrino $N$ is introduced,
but when more than
one exist cancellations are possible, including the natural case in which two
(nearly) degenerate Majorana neutrinos with opposite CP parities form a (quasi)
Dirac fermion. This is a general property: heavy (quasi) Dirac neutrinos
at the TeV scale or below are less
constrained and can reproduce light neutrino masses with less fine tuning.

We have reviewed the potential of various colliders to discover heavy neutrinos.
Their relative sensitivities strongly depend on the mass, mixing
and character of $N$. Some general statements can be made, however:
\begin{enumerate}
\item Lepton colliders have a better discovery potential for heavy neutrinos
with a significant coupling to the electron. They can be produced, up
to high masses, via $t$-channel diagrams.

\item LHC has the best discovery potential for a heavy Majorana neutrino with
$V_{eN} = 0$. The sensitivity of $e^- \gamma$ and $e^+ e^-$ colliders is
similar in this situation, but a factor of two smaller.

\item Neutrinos which couple to more than one charged lepton are easier to
detect, because: (a) LFV backgrounds are smaller in all
cases; (b) in $e^+ e^-$, $e^- p$ and $e^- \gamma$ collisions backgrounds
involving $\mu$ or $\tau$ leptons are also smaller.
However, the better observability
at lepton colliders does not translate into better bounds on $V_{\mu N}$ or
$V_{\tau N}$. Production cross sections are independent of these
couplings and, when $V_{\mu N}$ or $V_{\tau N}$ are sufficiently large
so that $\mu$ or $\tau$ final states dominate, decay branching
ratios are independent as well. This fact can be clearly observed
in Figs. \ref{fig:limits-ILC}, \ref{fig:limits}.

\item Dirac neutrinos are best studied in $e^+ e^-$ collisions, where they are
copiously produced and the environment is sufficiently clean.
\end{enumerate}

\begin{table}[h*]
\begin{center}
\begin{tabular}{ccccccccc}
& & & Majorana & & & & Dirac \\[1mm]
& ~ & Low & Intermediate & Large & ~~ & Low & Intermediate & Large \\
\hline
$e$ & & \begin{tabular}{c} $e^- \gamma$ \\ CLIC \\ ILC \\ $\mathrm{LEP} \otimes
\mathrm{LHC}$ \\ LHC \end{tabular} &
\begin{tabular}{c} CLIC \\ $\mathrm{LEP} \otimes \mathrm{LHC}$ \end{tabular} &
CLIC
& & \begin{tabular}{c} CLIC \\ ILC \\ LHC \end{tabular} & CLIC & CLIC \\
\hline
$\mu$ & &\begin{tabular}{c}  LHC \\ $e^- \gamma$ \\ ILC \end{tabular} &
 -- & --
& & LHC & -- & -- \\
\hline
$e,\mu$ & & \begin{tabular}{c} $e^- \gamma$ \\ CLIC \\ ILC \\
  $\mathrm{LEP} \otimes \mathrm{LHC}$  \\ LHC \end{tabular} &
\begin{tabular}{c} CLIC \\ $\mathrm{LEP} \otimes \mathrm{LHC}$ \end{tabular} &
CLIC
& & \begin{tabular}{c} CLIC \\ ILC \\ LHC \end{tabular} &
CLIC & CLIC \\
\end{tabular}
\caption{Summary of the relative discovery potential for heavy
Majorana and Dirac neutrinos, in the low ($100-400$ GeV), intermediate
($400-1000$ GeV) and large ($1-2.5$ TeV) mass regions, and coupling to $e$,
$\mu$ or both. In each cell, better to worse discovery potentials are ordered
from top to bottom. Dashes are shown when there is no significant sensitivity.
Results for mixing with the $\tau$ are analogous than for the muon.}
\label{tab:summ}   
\end{center}
\end{table}

Table~\ref{tab:summ} summarises the relative sensitivities of the different
colliders studied. It is useful to consider three approximate mass ranges for
heavy neutrinos: ``low'', from 100 to 400 GeV, ``intermediate'' from 400 GeV to
1 TeV and ``large'', from 1 to 2.5 TeV, which is close to the maximum
$m_N$ which can be probed at CLIC (see Fig.~\ref{fig:mass-coup}). It is also
convenient to take several limits
for the mixing of $N$: when it only couples to the electron, muon or tau, and
when it couples significantly to the electron and either muon or tau lepton.
In this latter case, we assume a significant coupling to the electron but with
decays dominated by muon / tau final states. We observe that for heavy Dirac
neutrinos, which naturally appear in some Little Higgs and extra-dimensional
models, $e^+ e^-$ collisions at CLIC and ILC provide the best limits.

\begin{table}[h*]
\begin{center}
\begin{tabular}{cccc}
& & Majorana  \\[1mm]
& Low & Intermediate & Large  \\
\hline
$e$ & $|V_{eN}| \leq 0.003-0.002$ &
      $|V_{eN}| \leq 0.002$ &
      $|V_{eN}| \leq 0.002-0.01$  \\[3mm]
$\mu$ & $|V_{\mu N}| \leq 0.022-0.1$ & -- & -- \\[3mm]
$\tau$ & $|V_{\tau N}| \lesssim 0.045-0.2$ & -- & -- \\[4mm]
$e,\mu$ & \begin{tabular}{c} $|V_{eN}| \lesssim 0.003-0.002$ \\
   $|V_{\mu N}| \lesssim 0.022-0.1$ \end{tabular} & 
   $|V_{eN}| \leq 0.001$ &
   $|V_{eN}| \leq 0.001-0.005$ \\[4mm]
$e,\tau$ & \begin{tabular}{c} $|V_{eN}| \lesssim 0.006-0.004$ \\ 
           $|V_{\tau N}| \lesssim 0.045-0.2$  \end{tabular} & 
           $|V_{eN}| \leq 0.002$ &
           $|V_{eN}| \leq 0.002-0.01$ \\[6mm]
& & Dirac \\[1mm]
& Low & Intermediate & Large \\
\hline
$e$ & $|V_{eN}| \leq 0.01-0.002$ &
      $|V_{eN}| \leq 0.002$ &
      $|V_{eN}| \leq 0.002-0.01$ \\[3mm]
$\mu$ & ? & -- & -- \\[3mm]
$\tau$ & ? & -- & -- \\[3mm]
$e,\mu$ & $|V_{eN}| \lesssim 0.01-0.002$ &
          $|V_{eN}| \leq 0.001$ &
          $|V_{eN}| \leq 0.001-0.005$ \\[3mm]
$e,\tau$ & $|V_{eN}| \lesssim 0.02-0.004$ &
           $|V_{eN}| \leq 0.002$ &
           $|V_{eN}| \leq 0.002-0.01$ \\[1mm]
\end{tabular}
\caption{Estimated 90\% CL upper bounds on the mixing of
heavy Majorana and Dirac neutrinos which couple to the charged leptons in the
left column, for the low ($100-400$ GeV), intermediate
($400-1000$ GeV) and large ($1-2.5$ TeV) mass regions.
Dashes are shown when there is no significant sensitivity.}
\label{tab:summ2}
\end{center}
\end{table}

We collect in Table~\ref{tab:summ2} the approximate 90\% CL bounds expected if
no heavy neutrino signal is found at any of the colliders discussed. They have
been obtained from the various limits and plots presented in section
\ref{sec:5}. Limits on $V_{\mu N}$ and $V_{\tau N}$ originate from $N$
production at LHC, and so they are relevant only for low $m_N$ values. They are 
shown for the Majorana case. Limits for a Dirac neutrino are not still
available nor can be safely estimated with present analyses. 
$5 \sigma$ discovery limits can be obtained from the figures presented by
multiplying by a factor $\sim 1.7-2$.

To conclude, we stress the importance of searches for heavy neutrinos at future
colliders. Light neutrino masses are at present the only piece of evidence for
physics beyond the SM, but their source is (and may remain forever)
unknown. A possible discovery in this direction would clarify the situation,
confirming and discarding possible scenarios for neutrino mass generation and
possibly leptogenesis.

\vspace{1cm}
\noindent
{\Large \bf Acknowledgements}

\vspace{0.4cm} \noindent
This review summarises and extends work presented at
``CP Violation and the Flavour Puzzle'', Symposium in Honour of Gustavo C.
Branco, GustavoFest 2005, Lisbon, Portugal, July 19--20, 2005;
XIX International Conference of Theoretical Physics ``Matter to the Deepest'',
Ustro\'n, Poland, September 8--14, 2005;
Corfu Summer Institute on Elementary Particle Physics, CORFU2005, Corfu, Greece,
September 4-26, 2005;
Euro GDR SUSY 2005 meeting, 2--5 November 2005, Barcelona, Spain; and
Third ECFA ILC Workshop, 14--17 November 2005, Vienna, Austria.
This work has been supported by MEC project FPA 2003-09298-C02-01,
Junta de Andaluc{\'\i}a projects FQM 101 and FQM 437 and
MIUR under contract 2004021808\_009.
J.A.A.-S. acknowledges support by a MEC Ram\'on y Cajal contract.


\end{document}